\documentclass{elsart}

\usepackage{amsmath,amssymb,amsfonts,epsfig, subfigure, float, setspace, lineno}

\psfigdriver{dvips}

\begin{document}

\begin{frontmatter}
\title{The species-area relationship and evolution}

\author{Daniel Lawson}, \author{Henrik Jeldtoft Jensen\corauthref{cor}}
\corauth[cor]{Corresponding Author. \ead{h.jensen@imperial.ac.uk}}
\address{Department of Mathematics, Imperial College London,  South Kensington Campus,  London, UK. SW7 2AZ}

\date{\today}

\begin{abstract}
Models relating to the Species-Area curve usually assume the existence of species, and are concerned mainly with ecological timescales.  We examine an individual-based model of co-evolution on a spatial lattice based on the Tangled Nature model in which species are emergent structures, and show that reproduction, mutation and dispersion by diffusion, with interaction via genotype space, produces power-law Species-Area Relations as observed in ecological measurements at medium scales.  We find that long-lasting co-evolutionary habitats form, allowing high diversity levels in a spatially homogenous system.
\end{abstract}

\end{frontmatter}

\linenumbers

\section{Introduction}

The number of species in a given region can be seen as a product of the evolutionary history of speciation, extinction and migration to that region.  Time variations in an ecology, whether induced by population dynamics or evolutionary dynamics, are caused by processes operating at the level of individuals; taxonomic structures, like species and genera, are emergent entities produced by the unceasing action of reproduction, mutation and annihilation of individuals. Hence it should be possible to derive the stability properties, abundance and distribution of species from a `microscopic' description in terms of dynamics at the level of individual organisms.  Such a framework must be able to act as a unified explanation of ecological structures such as the Species Area Relation (SAR) and  the Species Abundance Distribution (SAD) together with evolutionary aspects such as the temporal variation of the macroscopic averaged extinction rate and intermittency in the extinction events.  

The relationship between the number of species observed in an area and the area's size is one of the most basic questions in ecology but it is still the subject of much debate.  The number of species found in an area could increase with area size simply because more individuals are counted, and the form of this relation may be very different depending on the counting method used and details of the area \cite{SixSpecA}\cite{PossSpecA}.  For most measurement scales on non-island systems it seems that a power-law - $(diversity) \propto (area)^z$ - may be an accurate description, for the majority of fauna and flora types.  However, for other scales and for some data, other forms have been successfully fit \cite{SpecDiv}.  Here we consider those systems for which a power law provides a good fit - we will comment below on possible effects not included in our model which may be responsible for observed non-power law forms. 

Dynamical models typically {\em assume} the {\em existence} of a set of species as given structures classifying individuals. The dynamics at the individual level then determines how the assumed species are, say, populated and spatially distributed.  Particularly impressive examples of this type of models are Hubbell's \cite{NeutralTheory} 2001 `Unified Neutral Theory' and Durrett and Levin's \cite{DurrettLevinSAR} 1996 spatial voter model.  In the neutral models \cite{NeutralTheory}\cite{DurrettLevinSAR}\cite{NTecologyNature}\cite{ChaveNeutralReview} all individuals have the same birth, death and migration rate independent of which species they belong to.  Sol{\'e}, Alonso and McKane \cite{McKaneReview} have introduced a more general set of models in which an interaction matrix allow the assumed set of species to vary in their properties. Choosing specific forms for the interaction matrix reduces this model to a number of previously considered models - among these is Hubbell's neutral model. Realistic SAD and SAR are obtained from these models even in the case of neutrality between species. The SAD and the SAR has in addition been explained by an attractive geometric approach by Harte and co-workers \cite{HarteSelfSim}, who replaced dynamics by the assumption that the spatial distribution of species is self-similar and fractal; a prediction which has been confirmed from field data on birds in the Czech Republic \cite{finitearea}.  They concluded that a power law SAR was equivalent to a community level fractal distribution of species.

The Tangled Nature model (TaNa for short) is an attempt of developing a logically simple approach to evolutionary ecology.  From a few fundamental and generally accepted microscopic assumptions, macroscopic phenomenon such as macroevolution and ecological structures emerge.  The model is individual based with fluctuating population size, and the mutation prone reproduction occurs with probabilities determined by the interaction between co-evolving organisms.  The long time macroevolution in the model is consistent with observed temporal characteristics \cite{TaNaBasic}, the SAD compares well with observation \cite{TaNaNetwork} and most recently the model has been used to understand microbiological experiments on the relation between diversification and interaction \cite{TaNaEcoli}.  In the present paper we demonstrate how the Tangled Nature approach can be used to understand the SAR from an evolutionary individual based view point.

We will be introducing spatial aspects into the non-spatial TaNa model, in order to measure the SAR.  Essentially all good dispersion models produce reasonable fit with data (usually a power law) - e.g. the spatial models discussed above.  Power-laws are often observed in field data, but not universally \cite{Connor1979}, and we hope to eliminate two of the possible causes of the deviation - interactions and localisation (i.e. deme structure).  The interaction permitted in our model provides approximate power-law SAR regardless of strength, so inhomogeneity in migration or resource is a more likely source of observed deviations from power law in real systems, as such inhomogeneities are not considered here.

Here we consider species as dynamical quantities that emerge in genotype space.  We allow for spatial extension in a homogeneous physical environment, breaking the population into a number of spatial locations (with each species type forming separate demes) which in our model permits the construction of co-evolutionary habitats\footnote{We use the term co-evolution in the weak sense of species that have adapted due to interactions with other species.  We will also refer to these `co-evolutionary habitats' as simply `habitats' for brevity, as they are the only kind of habitat possible in our model.} of interacting species within each lattice point.  Individuals move by random dispersion as in the models mentioned in the previous paragraph.  The co-evolutionary habitats survive for very long time periods, during which local species abundances fluctuate around some average level.  Inside these habitats equivalence of individuals is observed, as a result of adaptation. The offspring probability of an individual depends on its genotype and on the composition of the local community in the local genotype space.  All individuals are subject to the same annihilation rate and only individuals that have evolved genotypes with an offspring probability that matches the killing probability are able to constitute species with a degree of temporal stability.  This leads to a certain degree of equivalence or neutrality to emerge amongst the dynamically generated species.  Since the offspring probability of an individual depends on the local occupation of genotype space, when individuals disperse to other habitats they usually do not have the same offspring probability as the members of the habitat they enter.  If species composition begins to change locally, then the entire habitat is usually affected, disrupting the local species composition.

Interaction allows for the extinction of well-established species on ecological timescales in the right invasion circumstances, giving realistic Species Abundance Curves (approximately log-normal \cite{TaNaBasic}).  Although species in the Tangled Nature model are dynamical and emergent, properties associated with random dispersal such as power-law SAR are observed.  The interaction allows distinct species to be separated in genotype space, in contrast with neutral models.  In hypercubic genotype space and in the absence of interaction species are clustered around a mean with separation occurring only by fluctuation and persisting for short timescales \cite{RechtsteinerNeutralGspace} (this also tested for the non-interacting version of our model, where the population essentially moves stochastically as one coherent cluster through genotype space).

The original Tangled Nature model defined by Christensen et al. \cite{TaNaBasic} has no spatial component, which we introduce by running copies of the model concurrently on a square lattice, allowing for interaction between lattice-points by migration.  The interaction between individuals at adjacent sites is therefore indirect, acting through genotype space only via the distribution of migrants, and the spatial aspect is discrete.  However, we can easily compare our results to that of the original model which has stability properties known to be close to observed systems \cite{TaNaTimeDep}\cite{TaNaQuasi}\cite{TaNaNetwork}.  The motivation for our approach is that genera that can move (animals and bacteria), or whose offspring can compete over distance for space (most plants) are modelled as locally well mixed, with spatial aspects considered on larger scales.

We begin with a recap on the non-spatial Tangled Nature model and its major features.  Then we detail our simple extension to the model introducing spatial dimensions using a square lattice of models. 

\section{Definition of the Model}
\label{SecDefinition}

We now define the Tangled Nature model.  We will be constructing the model on a periodic square lattice of length $X$.  Specific points on the lattice are referred to by their co-ordinates $(x,y)$.  Each point on the lattice may contain any number of individuals who, on any given time step, may migrate with probability $p_{move}$ to a neighbouring lattice point (our neighbourhood includes diagonals, and therefore is 8 lattice-points).  On each lattice point we run a TaNa summarised below and described in \cite{TaNaBasic}\cite{TaNaQuasi}, with interaction between lattice points via migration.  Each lattice point contains a number of species, made up of explicitely modelled individuals.  Similar approaches have been used many times, e.g. with each lattice point containing a local food web \cite{StaufferPatchFoodWebs}, or being used as the basic unit instead of individuals for models in which the two scales can be well separated (Gavrilets book \cite{FitnessLandscapes} considers this and many other situations).  Such separation of scales is not possible in our model, as the specifics of individuals control the invadability and stability of the local population.

The Tangled Nature model represents individuals as a vector $\mathbf{S}^{\alpha} = (S^\alpha_1,S^\alpha_2, ... , S^\alpha_L)$ in genotype space.  The $S^\alpha_i$ take the values $\pm 1$, and we use $L=20$ throughout.  Each $\mathbf{S}^{\alpha}$ string represents an entire species with unique, uncorrelated interactions, i.e. genotype space is coarse-grained.  The small value of $L$ is necessary for computational reasons as all genotypes exist \emph{in potentia} and have a designated interaction with all other possible organisms.  It is also possible to define the model slightly differently in terms of smooth traits, and correlate interactions over the trait space \cite{SimonsTaNa}.

We refer to individuals by Greek letters $\alpha, \beta, ... = 1,2,..., N(t)$ for a specific lattice point $(x,y)$.  Points in genotype space are referred to as $\mathbf{S}^a, \mathbf{S}^b, ...$, and many individuals (from any real-space location) may belong to a point in genotype space $\mathbf{S}^a$.  

In the TaNa model, all individuals are considered to die with equal probability $p_{kill}$, so it is most appropriate to systems where competition is for offspring space or resources (plants or bacteria, for example).  Only the probability to produce offspring is controlled by their interactions; however, the model is qualitatively the same regardless of whether varying killing or reproduction rates are used\cite{TaNaBasic}.  Reproduction occurs asexually, and on a successful reproduction attempt a daughter organism is produced which will be mutated with  probability $p_{mut}$.  When an individual $\alpha$ is chosen for processing, it will reproduce with probability:

\begin{equation}
p_{off}(\mathbf{S}^\alpha,t) = \frac{\exp[H(\mathbf{S}^\alpha, t)]}{1+\exp[H(\mathbf{S}^\alpha, t)]} \in (0,1)
\end{equation}

$p_{off}$ is defined in this way as it is the simplest way to translate $H(\mathbf{S}^\alpha, t)$ into a reproduction probability.  $H(\mathbf{S}^\alpha, t)$  is defined in Equation \ref{Hdef} and contains the bulk of the model, consisting of interaction and competition.  It is the average interaction (first term) and resource competition (second term) with all other individuals in the same spatial location.  Interactions are considered as an average (hence dividing by the population size $N(t)$) and we write it as a sum over all species rather than individuals, as individuals of the same species are identical.

\newfloat[H]{
\begin{equation}
\label{Hdef}
H(\mathbf{S}^\alpha,t) = \frac{1}{c N(t)} \cdot \sum_{\mathbf{S} \in S} J(\mathbf{S}^\alpha, \mathbf{S}) n(\mathbf{S},t) - \mu N(t)
\end{equation}
}

$c$ is a parameter controlling the interaction strength, $N(t)$ is the total number of individuals at time $t$ and  $n(\mathbf{S},t)$ is the number of individuals with genotype $\mathbf{S}$ at that point.  $\mu$ controls the carrying capacity of the system, preventing population growth when $N$ is of the order $1/\mu$.  The \emph{interaction matrix} $J(\mathbf{S}^\alpha, \mathbf{S})$ represents all possible couplings between all genotypes, each generated randomly in the range $(-1,1)$, being non-zero with probability $\Theta$.  Since the functional form of $J(\mathbf{S}^a,\mathbf{S}^b)$ does not affect the dynamics, provided that it is non-symmetric with mean $0$, we choose a form of the interaction matrix that speeds computation \cite{TaNaBasic}.  In the spatial version, we use the same $\mathbf{S}$ but allow the individuals to be located at a point in space, such that $\alpha = \alpha(x,y)$, $N=N(x,y,t)$ and $n=n(x,y,\mathbf{S},t)$.

Since the elements of $J$ are generated randomly, the pairwise interactions can be of the following types: mutualism (both positive), competition (both negative) and predator/prey (or parasitic) relations (one positive and one negative).  We do not allow for one-way interactions such as amensalism, apart from in the case where one interaction is randomly generated to be very small.  Also note that even in the case of extreme mutualism, resource is limited and competition will occur as the population increases, and so the negative term $\mu N(t)$ in Equation \ref{Hdef} is large.

The interactions modelled here are very general, though must occur through some medium which is not modelled explicitly.  For bacterial systems, this would be in the form of chemicals, meaning that the resource is modelled to some degree, but for plants it is more likely to be direct competition for offspring space.  The limiting resource, controlled by $\mu$ is different to any interaction facilitating resource, and might be space or a food source depending on the system under comparison.  There is only one `type' of resource, however, and as such we are only really modelling within a single trophic level, amongst individuals concerned with the same basic resource.  Thus, our model can be compared with data for herbivorous birds, or bacteria, or crop plants, but only for a predator-prey system when individuals on different trophic levels still compete for space.  This is not a problem for this papers purposes as most SAR data is drawn from a single family of species.  We are trying to model both the obvious food-web interactions as well as the multitude of perhaps weaker, hidden interactions.  It would be simple to add a number of additional resource types, with species drawing variously from different resources, but this adds a level of complexity unnecessary for the current questions.  This is instead considered as an extension to the model \cite{SimonsTaNa}.

In an offspring individual, each $\mathbf{S}^\alpha_i$ is mutated (flipped from 1 to -1, or from -1 to 1) with probability $p_{mut}$ from the parental $\mathbf{S}^\alpha_i$.  Thus mutations are equivalent to moving to an adjacent corner of the L-dimensional hypercube in genotype space, as discussed in \cite{TaNaBasic}.

A time-step consists of choosing a spatial lattice point with probability proportional to the population of the lattice point $N(x,y,t)$.  Then an individual\footnote{In previous versions a different individual was chosen for reproduction and killing actions.  Here we select only one individual and process it for reproduction, killing and movement for code efficiency reasons - above the level of fluctuations the two methods are equivalent.} $\alpha$ is chosen randomly from that lattice point. 

\begin{itemize}
\item
  $\alpha$ is allowed to reproduce with probability $p_{off}$.
\item
  $\alpha$ is killed with probability $p_{kill}$.
\item
  If the killing attempt was unsuccessful, $\alpha$ is moved to an adjacent lattice point with probability $p_{move}$.  Thus the effective $p^{eff}_{move} = (1-p_{kill})p_{move}$.
\end{itemize}

We define a generation as the amount of time for all individuals to have been killed, on average, once.  For a stable population size, this is also the time for all individuals to have reproduced once, on average.  Generations therefore are overlapping, and individuals have an exponential lifetime.  The choice of constant $p_{kill}$ does not appear to affect the general results - if we reversed the situation and allowed constant $p_{off}$ whilst varying fitness via $p_{kill}$, the same behaviour is observed (as the equilibria has $p_{off} \approx p_{kill}$ for all species).  We should therefore not observe results that are specific to either high infant mortality or high adult competition mortality, but we should observe features common to both competition types.

Although our model is asexual, we are operating on a sufficiently course-grained level that sexual reproduction can be considered as only possible between two individuals of the same genotype, and therefore is identical to the asexual case in our model, apart from when the abundance of a species is so low it would not be able to find a mate.  Whilst this permits comparison with data from both sexually and asexually reproducing species, this approximation will be invalid for many cases; we do not consider cross-over effects, for example.  One can think of our genotype space as modelling the genes that effect fitness, with `neutral' variation permitted in a type without being explicity modelled.  Some of the effects of sex could be incorporated into the mutation probability - others must simply be ignored.  We have not yet found any population level data that significantly contradicts our model, although clearly we miss a lot of the fine detail.  A discrepancy between our model and observed data which is only present for sexual species could shed light on the population level effects of sexual reproduction.

\section{Behaviour of the model}

We will first review the behaviour of an isolated system, and then use this to help interpret the results on an $X$ by $X$ square lattice with periodic boundary conditions.

Unless otherwise stated, the parameters used will be: $\Theta=0.25$, $c=0.05$, $\mu=0.05$, $p_{mut}=0.01$ and $p_{kill}=0.1$; see \cite{TaNaBasic} for more details.  These are chosen to keep the population of the entire system from exceeding about $30000$, keeping computation to reasonable levels and allowing for averaging.  The population of a specific lattice point is low compared with previous studies (around $300$ in this study), increasing the strength of stochastic effects - hence the other parameters are chosen to cancel out this effect to some degree.  Although the mutation rate is unrealistically high, it still reproduces the correct qualitative effects found in real systems \cite{TaNaQuasi}, and simply gives a higher turnover of quasi-evolutionary stable strategies as defined in Section \ref{SecIsolated}.  It should be stressed that the qualitative behaviour observed here is seen at mutation rates down to $10^{-8}$.

\subsection{The isolated TaNa model}
\label{SecIsolated}

We briefly review the behaviour of a single TaNa model as given by \cite{TaNaBasic}\cite{TaNaTimeDep}.  The model exhibits a number of quasi-evolutionary stable strategies (q-ESSs) in which the frequency distribution in genotype space remains constant (with some small fluctuations); these q-ESSs are also observed in differential equation style models \cite{LargeSpeciesShifts}\cite{TokitaStableNetwork}.  The q-ESSs are named after the Evolutionary Stable Strategies \cite{MaynardSmithGT}, or ESSs, found in game theory.  If we think of competing individuals which may adopt a strategy for survival, then an ESS is a strategy which, if adopted by the entire population, will not be invadable by any other strategy.  The strategy of an individual defines its actions in all circumstances; in our model the strategy is the list of interactions with all other types.  It is the strategy of the population as a whole that is important here, given by the proportion of individuals having each individual strategy.  A stable strategy is thus a set of individuals who cannot be invaded by an increase in any of the other types (that is, if one type gains population, it loses interactions and therefore will lose population).  However, because we include mutation, the strategy must also be stable to an influx of all local types.  This list of local types is only a \emph{subset} of all possible invader strategies, and so a population may be \emph{quasi}-stable; that is, stable to all local mutations but not to distant genotypes which can only be reached by stochastic fluctuations (as they are separated from the population by a fitness minima).  These distant genotypes can do well in the q-ESS population, and therefore destabilise it as their population grows.

\begin{figure*}[ht]
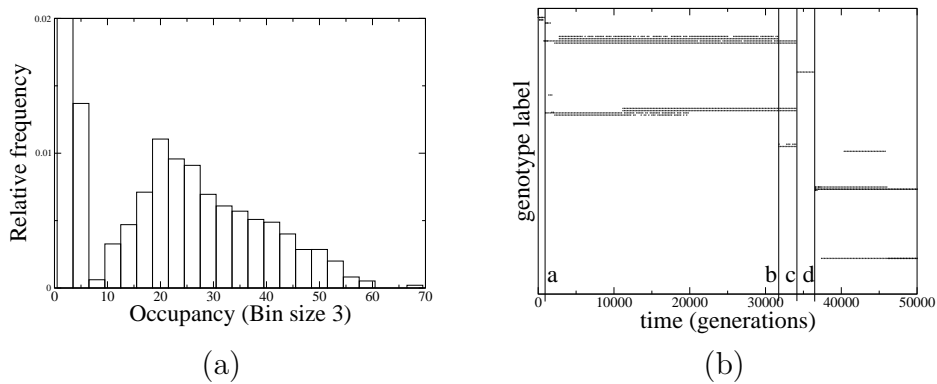

\begin{tabular}{cc}
  \begin{minipage}{.45\textwidth}
    \centering
    \epsfig{file=FigGenoOcc.eps,width=0.9\textwidth}
  \end{minipage}
  &
  \begin{minipage}{.45\textwidth}
    \centering
    \epsfig{file=FigExamplePlot.eps,width=0.9\textwidth}
  \end{minipage}
  \\
  \begin{minipage}{.45\textwidth}
    \centering
    (a)
  \end{minipage}
  &
  \begin{minipage}{.45\textwidth}
    \centering
    (b)
  \end{minipage}
\end{tabular}
\caption{\scriptsize{(a) Species abundance, or relative occupancy of points in genotype space, averaged over all 50000 generations and 380 runs.  There is a distinct difference between those genotypes with occupancy less than about 6 and those with an occupancy greater than 16, with only a very small amount in between, most of which come from transition period species. (b) An example occupancy plot showing all species with occupancy $n_a>8$ at time $t$ as a dot.  Species are not meaningfully ordered. q-ESS periods are shown as horizontal lines, with most transition periods (apart from the very slow one from around time $11000$ to $20000$) too short to see on this scale.}\label{FigGenoOcc}}
\end{figure*}

We operate with parameters that give a reasonable number of q-ESS switches within the first 50000 generations - most of the work analysing this region was done in \cite{TaNaQuasi}.  During these q-ESSs (shown in Figure \ref{FigGenoOcc} (b)), a number of genotypes (the `wildtypes') are highly occupied - other genotypes are only present due to mutation from the wildtypes, and are frequently eliminated by stochastic events (see Figure \ref{FigGenoOcc} (a)).  As our genotype space is coarse-grained, these `sub-species' do not inherit interaction properties from a wildtype - despite this, a natural species-concept emerges as a simple result of interaction in a genotype space.  Thus our diversity measure is the `wildtype diversity': simply the number of wildtypes in the system.  Wildtypes are defined as genotypes with occupancy of eight or greater (a definition which is valid only for these parameter ranges).  We have tested other diversity measures such as the Shannon-Wiener Index and our results are qualitatively the same regardless of measure used, but these are primarily designed to avoid sampling problems \cite{MeasBioDiv} and so are less relevant to computer simulations.

In \cite{TaNaTimeDep}, it is shown that the average q-ESS length increases with time, due to increasing stability in the network of active interactions, increased population size and hence increased diversity (as larger populations are more likely to be stable to stochastic fluctuations, and q-ESS interactions tend to be positive).  Note that these effects occur only \emph{on average} - it is possible for the system to move to a less stable, smaller population after a disordered phase, and it is also not always true that higher populations are more stable (or more diverse), just that they are on average.

During the q-ESS, wildtype occupation fluctuates around some constant level, and sub-species appear and dissappear by mutation, without affecting the stability of the q-ESS state. Biologically, a q-ESS has all species in a q-ESS occupying a fitness maxima (that is, all mutations have lower fitness - fitness meaning offspring probability in this case), which the system has found during a transition.  Each species in the q-ESS must have reached a population equilibrium, so that $p_{off} \approx p_{kill}$, and all mutants from each species must have $p_{off} < p_{kill}$ when their own population is low.  This is easier to achieve for a low diversity, but when a stable state is found at higher diversities, the chance that an invader will destabilise the q-ESS is lower as invaders will be at significantly lower fitness on average (due to the increase in the average population $N$ from those positive interactions).  It is therefore of interest to analyse the transition more closely, in order to understand why the q-ESS forms in the way it does.

Transitions appear in many forms, depending on the configuration of the genotype space surrounding the wildtypes.  There are two events that can force a q-ESS to end:

\begin{itemize}
\item
If a genotype with $p_{off} > p_{kill}$ can be reached, then there will be a period where the mutant population is still vulnerable to accidental extinction, followed by an exponential growth period if the mutant population grows large enough.  This will usually quickly upset the configuration of the local population, leading to transition.
\item
If one of the wildtype species had low average population then it can become accidentally extinct.  In some cases other species will not depend on this species and the system enters a similar q-ESS with reduced diversity; in other cases, the stability of the q-ESS is upset and a transition occurs.
\end{itemize}

Once the system enters a transition, one of the following may happen:
\begin{itemize}
\item
The disruption is minor and the system remains stable with a new q-ESS configuration.  The transition period is not well defined in this case.
\item
Wildtype species no longer all have $p_{off} = p_{kill}$.  The populations will change in order to regain this relation.  It is possible that a species may become extinct, leading to stage 2 above.
\item
One of the low population mutant species in the system will gain $p_{off} > p_{kill}$ and so will enter phase 1 above.
\end{itemize}

Clearly, this is an iterative process and can last for a very long time - forever if $c$ or $p_{mut}$ are very large, so pushing the system past the `error threshold' \cite{TaNaQuasi}.  It is additionally complicated because these processes are all really running simultaneously, and responding to each other.  What is clear, though, is that there is always favoured species in the system, and from simulations we see that the number of favoured species does not change significantly from q-ESS periods.  In \cite{TaNaBasic} it is shown that transition periods retain the distinction between (short lived in this case) wildtypes and mutants, resulting in a very similar (possibly identical) SAD.  Since the transition periods are very short, any deviation from the q-ESS SAD is negligible and for an instantaneous observation they are indistinguishable (as stochastic noise is high).  Transitions also provide a way for a species to mutate to a distantly related genotype quickly.  Because there is a high interaction  between all types, and the number of types is often quite high, most configurations are not q-ESS.  It is therefore unlikely that the initial invaders of a q-ESS will be successful in the long term - they instead will be in turn invaded by a second set of mutants.  This process continues until a q-ESS is found, and so there is an effective selection gradient away from the wildtypes during this time, leading to very large and fast changes in genotype acting for short periods of time.

The species abundances are of log-normal form as observed in many real systems \cite{TaNaNetwork} provided that the interaction probability $\Theta$ is high, as in the cases we consider, and the lifetime distribution for species is wide-tailed as in real data \cite{TaNaBasic} (following a power-law).  More details on the network properties of the Tangled Nature model is available from \cite{TaNaNetwork}, and an in-depth analysis of the time dependence of many of the observables such as diversity and total population is presented in \cite{TaNaTimeDep}.  Similar work by Zia and Rikvold \cite{RikvoldTanaBasic}\cite{RikvoldTaNaFluctuations} deals with a simplification of the non-spatial case.  In both models the q-ESS wildtypes are characterised as different to transition period wildtypes because their mutants do not interact favourably with the q-ESS population, and so are suppressed.

\subsection{The Tangled Nature Model on a spatial lattice}

We now introduce a square spatial grid of length $X$, each containing a TaNa model, and allow the lattice-points 
to interact by migration; migration probability refers to the chance of moving to \emph{any} neighbouring site, 
chosen randomly from the 8 nearest neighbours, and we assume a periodic boundary.  Just this simple addition to 
the basic TaNa model gives rise to naturally occurring Species-Area Relations, or SARs.

Unlike the non-spatial version of the model, initial conditions are relevant.  All possible starting configurations 
reduce to one of the following two initial conditions: 

\begin{enumerate}
\item
Individuals are generated with a random genotype and placed on a random lattice point until the total starting population is reached.
\item
A single lattice point is allowed to evolve as a separate system until a q-ESS is formed.  This q-ESS is copied 
to all other lattice points to give a quasi-stable, identical initial starting condition at all points.
\end{enumerate}

\begin{figure}
  [htb]
  \begin{center}
    \epsfig{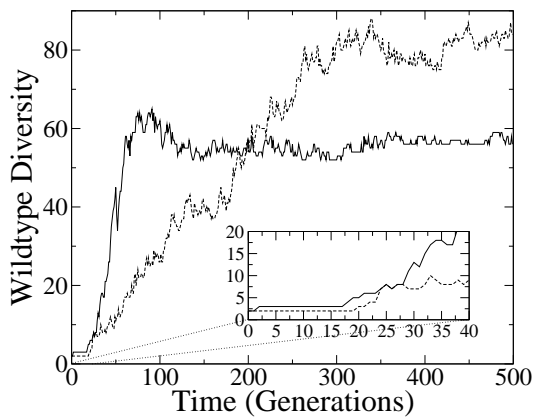}
    \caption{\scriptsize{Wildtype diversity against time for 2 initial systems consisting of the same stable q-ESS at all lattice points (initial condition type 2).  Diversity remains constant for around 20 generations, after which an increase is seen.  In one run (solid line), the increase occurs very rapidly but in the other (dashed line) the increase is more gradual yet reaches higher levels.  Once a stable level (on ecological timescales) of high diversity is found, the evolutionary dynamics occurs in the same manner as initial condition type 2, random seeding.} \label{FigDivVSTimeInitial}}
  \end{center}
\end{figure}

Procedure 2 represents the biological case where a small species set is exposed to a larger spatial range, and so colonises it.  The initial q-ESS used in procedure 2 has stability properties that can differ greatly - see Figure \ref{FigDivVSTimeInitial}.  It can vary in absolute stability (how long it will last for), but spatial duplication means that the number of stable q-ESSs that can be found from the initial transition is relevant, as this controls how quickly diversity will increase when a transition does occur in the system.  Procedure 2 therefore introduces a high stochastic variation resulting in a (sometimes sharp, sometimes smooth) diversity increase after an initial (possibly very long) wait.

Procedure 1 bears some resemblance to the colonisation of a new area of land by many species simultaneously.  It results in an initially high diversity as different q-ESS states form at all points.  This decreases quickly to an similar level found from procedure 2.  However, after this time, the two procedures are equivalent; hence in our analysis we shall consider only initial random seeding, i.e. procedure 1, in order to standarize the initial diversity level.  We then allow the system to evolve for a long time ($40000$ generations) before observation to allow an ecology to form.

The introduction of space has many implications for the model.  In the non-spatial case, there were two timescales: the average lifetime of an individual, and the average lifetime of a q-ESS, which increased slowly with time.  In the spatial case, we have a third timescale: the time taken for information of a transition to be transmitted to the other side of the system.  As this occurs only through transitions at all intermediate lattice points, this can be very long, much longer than the simulation time.  Another complication is that average q-ESS lifetime now depends strongly on the state of neighbouring lattice points, as migrants from different q-ESSs are disruptive but migrants from similar q-ESSs can actually stabilise a possible transition.  Thus, time averaging is not possible in large systems, and collecting data on the SAD becomes very difficult.  We therefore focus on calculating the SAR: that is, the relationship between the number of species found in an area and the size of the area.  We distinguish between the two size measures: the \emph{scale} as the sub-area measurement of a system with \emph{size} $X$.

SARs come in many forms, depending on the measuring system used.  Specifically, quoting \cite{SixSpecA}, there are 3 main properties : ``(1) the pattern of quadrats or areas sampled (nested, contiguous, noncontiguous, or island); (2) whether successively larger areas are constructed in a spatially explicit fashion or not; and (3) whether the curve is constructed from single values or mean values''.  We obtain nested, successive, mean value data.  Thus, for all scales, measurement squares are contained within a larger scales' measurement square, no shapes other than square are considered and we are averaging over all possible measuring squares from a specific scale. \cite{SixSpecA} and \cite{PossSpecA} discuss the implications for this.

Approximate SAR power-laws are often encountered in real systems at `medium' scales: that is, for areas that are smaller than the continent/land-mass that they are found on, but large enough to obtain a reasonable sample.  Good examples are plant species in Surrey, UK, (\cite{SpecDiv}, page 9) or bird species in the Czech Republic \cite{finitearea}.  When looking at other scales different SARs can be obtained; the distinction between scales is one that varies with environment and habitat types, and many functional forms of SAR can be found somewhere.  A general rule (p277 of \cite{SpecDiv}) is that inter-provincial relations follow power-law SARs with exponent larger than intra-provincially; islands inside a province will also have a larger exponent than the whole province itself (thus having smaller diversities).  A single run in our model corresponds to a single isolated province as it is spatially homogenous and self-contained.

A specific instance of our model will not have any real world equivalent, as we have selected genotype space interactions and our initial position in it randomly.  However, averages over our model should correspond to (large and thus self averaging) real systems for which our assumptions are approximately valid, as we are effectively averaging over the possible realisations of genotype space.  Any real world system that does not conform to this average will be affected by an effect not modelled here - for example, the geography or resource distribution may be an important factor.

\begin{figure*}[htb]
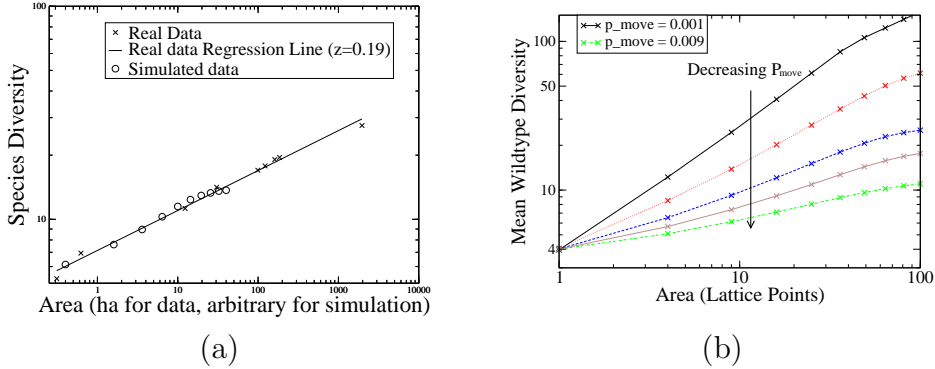

\begin{tabular}{cc}
  \begin{minipage}{.45\textwidth}
    \centering
    \epsfig{file=FigrealSA.eps,width=0.9\textwidth}
  \end{minipage}
  &
  \begin{minipage}{.45\textwidth}
    \centering
    \epsfig{file=FigSAR_varyingPmove.eps,width=0.9\textwidth}
  \end{minipage}
  \\
  \begin{minipage}{.45\textwidth}
    \centering
    (a)
  \end{minipage}
  &
  \begin{minipage}{.45\textwidth}
    \centering
    (b)
  \end{minipage}
\end{tabular}
\caption{ \scriptsize{(a) SAR Data for Hertfordshire plants taken from \cite{SpecDiv}(Fig 2.2) plotted with simulated data, assuming 1 lattice-point is a 0.4ha plot ($p_{move}=0.025$) evolved for 40000 generations.  (b)  Simulated, evolved SAR plotted for varying $p_{move}$ from $0.001$ to $0.009$ (in steps of $0.002$); the shape and start point remains the same, with only the exponent changing.} \label{FigSAmeasures}}
\end{figure*}

Real systems have $z$-values between 0.15 and 0.4 depending on the details of the system \cite{SpecDiv}.  Figure 
\ref{FigSAmeasures} illustrates real SAR data from Hertfordshire plants and shows a sample simulation SAR.  Both 
describe a power-law as are they are linear in log-log space, $\log S = z \log A + \log \alpha$, hence the slope 
of this line (the \emph{z-value}) is the major controlling factor in how quickly diversity grows with area.  For 
example purposes, we have chosen the area of a lattice-point arbitrarily as 0.4ha.  However the true size of a 
lattice-point in our model is not well defined as the TaNa model implicitly assumes all species are of equal 
spatial extension.  Hence we are now concerned only with the scaling relation: the form of the SAR being close 
to a power-law and the value of the exponent in that power-law.

As each run is a separate instance with its own evolutionary history, the diversity and $z$-value variation between 
runs is high unless the size is much larger than the species range; however, the power-law rule holds for all instances.

\begin{figure*}[htb]
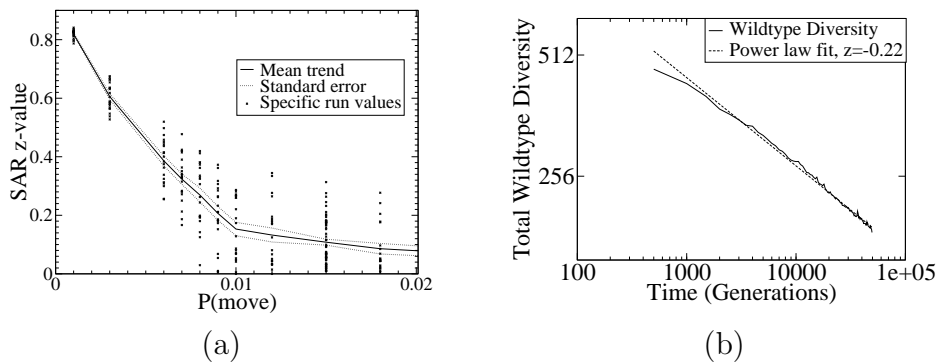

\begin{tabular}{cc}
  \begin{minipage}{.45\textwidth}
    \centering
    \epsfig{file=FigZtrend.eps,width=0.9\textwidth}
  \end{minipage}
  &
  \begin{minipage}{.45\textwidth}
    \centering
    \epsfig{file=FigDivVStime.eps,width=0.9\textwidth}
  \end{minipage}
  \\
  \begin{minipage}{.45\textwidth}
    \centering
    (a)
  \end{minipage}
  &
  \begin{minipage}{.45\textwidth}
    \centering
    (b)
  \end{minipage}
\end{tabular}
\caption{ \scriptsize{(a) z-value calculated from the wildtype diversity evaluated between 40000 and 50000 generations, 
showing individual z-values from runs (on a 10x10 lattice).  Note the two distinct regions - $p_{move} < 0.01$ where
 species do not spread large enough distances for finite size effects to matter, and $p_{move} > 0.01$ where in some 
 runs, species can span the entire system.  (b) log-log plot of diversity as a function of time for a 20x20 system 
 with $p_{move}=0.005$.} \label{FigZtrend}}
\end{figure*}

The simulated data in Figure \ref{FigSAmeasures} has a slightly reduced tail from the expected power-law values, due to the finite area of the simulation.  By holding a fixed system size ($X=10$ is chosen as be the maximum we can simulate with sufficient averaging ability) and varying $p_{move}$ (Figure \ref{FigZtrend} (a)) we can understand these cutoffs more fully.

Figure \ref{FigZtrend}(a) shows the individual values of $z$ for varying values of $p_{move}$ together with the average.  The values are distributed about some mean, which decreases approximately linearly with increasing $p_{move}$ for $p_{move} < 0.01$.  however, above $p_{move} = 0.01$ we observe that some of the runs give a near-zero $z$-value, i.e. a constant SAR curve, meaning that species are spanning the system.  The correlation length of the system has reached the system size and boundary affects will irrevocably effect the results.  With increasing $p_{move}$ the average patch size of each q-ESS increases, and thus the probability of finding a patch the size of the system increases.  In non-evolutionary models, one can avoid this problem by considering migration from a `pool' of constant species makeup \cite{IslandBiogeog} but in evolving systems the pool must be modelled explicitly.

Figure \ref{FigZtrend}(b) shows the time dependence of diversity.  Although new species are produced at all times, and new q-ESS states can be formed, they do not seem to do so at a rate that matches diversity loss.  The time taken to reach a single q-ESS state diverges with area, taking of the order $10^{12}$ generations for a single q-ESS to be reached for a 20x20 system, or $10^9$ generations for a 10x10 system.  As diversity can increase drastically at any time if a single species can destabilise the dominant q-ESS, it is unlikely this would not continue forever.  Instead, we would effectively be restarting the system with a procedure 2 initial condition; however, the stability of this highly evolved q-ESS is much higher than a random q-ESS taken from initial conditions, and so the time taken to see a restarted system may be very long (as q-ESS lengths are power-law distributed, this time has mean infinity - however, it does occur eventually, as there is no truly stable state in this model).

\begin{figure}
  [htb]
  \begin{center}
    \epsfig{file=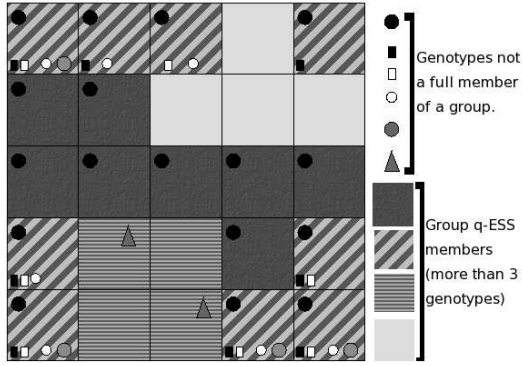,width=0.5\textwidth}
    \caption{\scriptsize{Spatial distribution of species on a small (5x5) periodic lattice after 50000 
    generations, with background shading for each point representing the basic q-ESS members and symbols 
    representing all genotypes that do not completely fit into a q-ESS category.  Some of these genotypes 
    are active in more than one q-ESS state (e.g. black circle) and others operate in subsets of a specific 
    q-ESS state (e.g. grey triangle).  All species are located in contiguous lattice-points, and it is 
    possible for some patches to span the entire area.} \label{FigSpatialDistrib}}
  \end{center}
\end{figure}

In the Spatial TaNa model, illustrated in Figure \ref{FigSpatialDistrib}, the spatial distribution of species is 
confined to a contiguous patch.  Non-contiguous patches seem to be rare as patches are more easily invaded at 
patch corners due to the positive self-reinforcement of a q-ESS type in the centre.  Species will generally coexist 
with a specific set of other species, forming fairly distinct q-ESS states of 3-8 species (shaded regions).  
However, there are many cases where the majority of q-ESS members remain constant but one species is swapped 
out for another.  Thus in some cases there is a smooth transition spatially between one q-ESS type and a completely 
different q-ESS type, with many transients along the way containing subsets of each (e.g. dense forest fading to 
woodlands then to grassland).  In other cases the coexistence is more essential and there will be a distinct line 
between one species set and another.

In toroid geometry, any observations of greater than half the total size are affected unaccountably by 
the periodic boundary so we restrict conclusions from scales less than $X/2$, which do appear to be 
truly power law related (tested for up to $X=20$).  Unfortunately, this size restriction does not permit 
the testing of self-similarity by any other means than the power-law relation, and we cannot tell if 
non-contiguous patches patches might form in larger simulations.  It is possible that species distribution is truly 
self-similar in our model, whether the patches are or not, as species may survive in several different patches.

We can also consider this system in the absence of mutation, so considering a `population dynamics' version of the model.  Here, initial conditions are very important as no new species can ever be added.  The quasi-stability observed previously will also change nature as the only possible disturbance is migrant species.  If we for the moment consider a single lattice site with randomly chosen species, the behaviour is similar to the usual case with mutation in that the number of species condenses down to a small number which are mutually stable.  As there can be no invasion, the only pressure is accidental death.  This occurs with very low probability for moderate population numbers as the form of $p_{off}$ ensures that there is a restoring pressure to the equilibrium.  The system will always find a steady state (which, rarely, may have only one species in if the species that survived the low population stage happen to all have non-mutualistic interactions).

\begin{figure}
  [htb]
  \begin{center}
    \epsfig{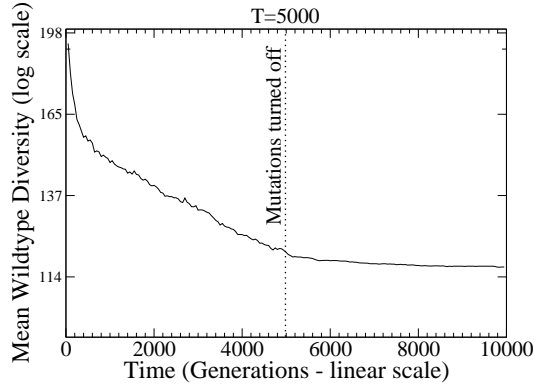}
    \caption{\scriptsize{Time dependence of diversity: for the first 5000 generations, mutations are permitted ($p_{mut}=0.001$), and are then stopped (averaged over 20 runs).  The system decay rate decreases markedly, but still follows a power law.} \label{FigPopDynamics}}
  \end{center}
\end{figure}

However, on a spatial lattice things are different.  If we choose to evolve a q-ESS to copy to all points then clearly the system will contain only this q-ESS forever, as there is no source of change.  If we start the system with random individuals, however, then the initial states found in each lattice point will be very different and so migrants may have significant impact.  In this case, we see a relaxation in diversity of similar form (power law) to the mutation case.  However, the rate of decay (the exponent for the decrease of diversity with time) is smaller compared to the evolving case.  A species area relation of the same form as in the evolving case is still seen, complete with slight S shape form.  If we start with an evolved system with a reasonable SAR, and then turn off evolution, we see that the decay with time of the diversity decreases drastically, as the system almost `freezes' (Figure \ref{FigPopDynamics}).  The SAR form will not change drastically, but the exponent will continue to decrease very slowly as the number of species, and the number of distinct q-ESS decreases.

This behaviour shows that it is population dynamics that give the SAR power law form, and that our formalism does not permit mutations to spread through the system with sufficient speed to offset extinctions.  Instead, evolution permits the generation of `better' q-ESS that can spread through the system more quickly, accelerating the rate of species loss.  However, evolution is required to produce diversity in the first place, and allows it to spread very quickly throughout the system as seen in Figure \ref{FigDivVSTimeInitial}.  In our model, environmental factors (changing in space and/or evolutionary time) are necessary for preventing the collapse of the SAR once it is formed.

\section{Discussion}

Our SAR results bear striking similarity with those of a neutral `voting' model of Durrett and Levin \cite{DurrettLevinSAR}.  The form of the SAR in both is almost power-law, with a slight s-shape produced by boundary effects.  They find that the z-value decreases with decreasing speciation rate (which is equivalent to immigration rate, if new species are introduced from another land mass, for example).  In our model with interactions and explicit genotype space, we find that z-value decreases with increasing migration rate inside the system.  Mutation occurs at constant speed, so increasing migration rate, e.g. Figure \ref{FigZtrend}(a), \emph{decreases} the relative spread of a new species, instead causing transitions to an already existing q-ESS and so reinforcing currently existing species.

Essentially, internal migration rate reduces the relative effect of mutations, and so produces the inverse effect of the immigration rate of new species from outside the system (which is equivalent to mutation in a point-mutation representation without consideration of genetics).  High mobility (i.e. migration and immigration rates) for a family of species mean better mixing and so less chance for spatial segregation of species within a single family - the standard explanation for why birds generally have lower z-values than land species.  Conversely, e.g. on islands, it allows species from elsewhere to arrive, so possibly increasing diversity (as argued in \cite{DurrettLevinSAR}).  Which effect dominates will depend on the geography in question - i.e. the size of the local groups of individuals, and the separation between them.  A more detailed model is required to probe this more fully.

Magurran and Henderson \cite{MagurranNature}, noted that permanent fish species have log-normal SAD whilst transient species have a log-series distribution.  Our local q-ESS has the same distribution, with a log-normal like distribution for the wildtypes and a log-series like for mutants and migrants.  For low mutation rates and high migration rates, clearly migrants will outnumber local mutants and we will observe the exact same distribution near the q-ESS patch borders.  Here, the distinction between the two types is of fitness - the wildtypes with a log-normal like SAD are all equally fit in that they have a reproduction rate exactly balancing the death rate; the migrants with a log-series like SAD are all less fit and rely on repopulation from an external pool.

The Tangled Nature model on a spatial lattice reproduces many of the observed features in real systems without making any \emph{a-priori} assumptions about the existence of species. Instead, species and their spatial distributions are allowed to form naturally by co-evolution from simple rules applied only to individuals.  Unfortunately, the model is currently too computer intensive to allow simulation of the very large scales (and higher migration rates) expected in real systems.  However, a near power law is clearly produced as a simple result of species forming patches of many sizes, themselves the product of diffusive dispersion with reproduction and mutation when local interaction is permitted.  Mutation is necessary to give `raw material' for new species to be formed.

Co-evolutionary forces are sufficient to allow (co-evolutionary) habitat differentiation (as shown in the co-habitation of competing E.coli strains in \cite{DiverseEcoli}), and the number of different habitats increases with area as a power-law.  Thus power-law SARs are observed, as the number of habitats can drive the diversity increase with area \cite{SpecDiv}, and these persist over long timescales and in the absence of geographical differences.  The evolutionary history therefore relates to the production, and $z$-value, of power-law like SARs and may be important in many cases \cite{SpecDiv}.  

The habitat differentiation produced by co-evolution allows species to be locally equivalent whilst interacting strongly, and maintains differences in offspring probabilities when removed from its favoured habitat.  Thus we find equivalence whenever individuals have had time to adapt to the homogeneous killing probability, which corresponds to a situation where individuals die mainly due to some more our less species independent stochastic killing mechanism.  An example of such a system might be `climax' stage of forest succession \cite{Biology}\cite{EvEcol}, where species makeup is approximately constant (over a sufficiently large area and time average) and the ratio of births to deaths are close to unity for all species.  Species measured in the field that were found to be non-equivalent \cite{ChaveNeutralReview} may be considered in the context of Tangled Nature to be transitionary, or may  simply be out of the habitat they were originally adapted to - the equivalence predicted in our system is very local, but can be formed over distances by the correct migration composition of species.

Individuals from species not found locally are generally poorly adapted to the local environment and go quickly extinct.  Rarely, however, species with $p_{off} > p_{kill}$ can invade and their increased chance of survival over the general population allows the species to flourish initially - providing a method for fast speciation from an initial mutant.  In addition, during transitions, intermediate genotypes are successful which may be replaced by other genotypes before a q-ESS is established, overcoming the `fitness barrier' to distant genotypes, with \emph{all intermediates occupying fitness maxima}.  Thus, speciation can occur quickly, and to species distantly related.  This contrasts the `fitness landscape' viewpoint (For a review, see e.g. \cite{DrosselReview}), in which speciation requires passing through a fitness minima.  It also solves a problem seen in neutral theories, which require external pressure such as allopatric speciation (i.e. isolating a whole community for mutation by ``random fission'' \cite{HubbellRicklefs}\cite{RicklefsNeutrality}, instead of using the traditional point mutation used here and in much of the literature) if realistically fast speciation and extinctions are to occur \cite{ChaveNeutralReview}.

We have identified the stability of species, fast extinctions and separation in genotype space as the main differences between our interacting model and neutral models.  The wildtypes in our system are locally equivalent, and it is the patches of these wildtypes that are producing the power-law SARs observed.  Wildtypes are thus equivalent most of the time but not when found outside their own habitat, where they suffer a reproductive disadvantage.  This is consistent with the non-neutrality observed in nature and may explain why neutral dynamics do so well at predicting SARs and SADs.  The non-neutrality is only important during transitions (which, in the spatial model are usually local events), but the number and distribution of species does not change, only the specific type of species.  These effects cannot be observed in instantaneous measures, or in time averages.

The spatial Tangled Nature model provides a simple general framework containing the basic properties of diffusive dispersion, reproduction and mutation on the level of individuals, it allows taxonomic structures to emerge and produces a large number of observed macroscopic ecological phenomenon - species abundance, long-lived species, fast extinctions, power-law lifetimes, intermittent dynamics, and, as demonstrated in the present paper, species-area relations.

\section*{\small{Acknowledgements}}
\scriptsize{We thank Andy Thomas and Gunnar Pruessner for providing assistance with processing the model 
and setting up the BSD cluster, speeding computation enormously.  We also thank the Engineering and Physical 
Sciences Research Council (EPSRC) for Daniel Lawson's PhD studentship.}

\nocite{PurvesPacala}

\let\oldbibliography\thebibliography
\renewcommand{\thebibliography}[1]{%
  \oldbibliography{#1}%
  \setlength{\itemsep}{0pt}%
}

\begin{spacing}{0}
\bibliographystyle{unsrt}
\begin{scriptsize}
\bibliography{evolution}

\begin{thebibliography}{10}

\bibitem{SixSpecA}
Samuel~M. Scheiner.
\newblock Six types of species-area curves.
\newblock {\em Glob. Ecol. \& Biogeog.}, 12:441--447, 2003.

\bibitem{PossSpecA}
Even~Tj\o rve.
\newblock Shapes and functions of species-area curves: a review of possible
  models.
\newblock {\em J. of Biogeog.}, 30:827--835, 2003.

\bibitem{SpecDiv}
Michael~L. Rosenzweig.
\newblock {\em Species diversity in space and time}.
\newblock Cambridge University Press, The Edinburgh Building, Cambridge CB2
  2RU, 1995.

\bibitem{NeutralTheory}
Stephen Hubbell.
\newblock {\em The Unified Neutral Theory of Biodiversity and Biogeography}.
\newblock Princeton University Press, 41 William Street, Princeton, New Jersey
  08540, 2001.

\bibitem{DurrettLevinSAR}
Rick Durrett and Simon Levin.
\newblock Spatial models for species-area curves.
\newblock {\em J. theor. Biol.}, 179:119--127, 1996.

\bibitem{NTecologyNature}
Igor Volkov, Jayanth~R. Banavar, Stephen~P. Hubbell, and Amos Marita.
\newblock Neutral theory and relative species abundance in ecology.
\newblock {\em Nature}, 424:1035--1037, 2003.

\bibitem{ChaveNeutralReview}
J.~Chave.
\newblock Neutral theory and community ecology.
\newblock {\em Ecol. Let.}, 7:241--253, 2004.

\bibitem{McKaneReview}
Ricard~V. Sol\'{e}, David Alonso, and Alan McKane.
\newblock Self-organised instability in complex ecosystems.
\newblock {\em Phil. Trans. R. Soc. Lond. B}, 357:667--681, 2002.

\bibitem{HarteSelfSim}
John Harte, Tim Blackburn, and Annette Ostling.
\newblock Self-similarity and the relationship between abundance and range
  size.
\newblock {\em Am. Nat.}, 157:374--386, 2001.

\bibitem{finitearea}
Arno\v{s}t~L. \v{S}izling and David Storch.
\newblock Power-law species-area relationships and self-similar species
  distributions within finite areas.
\newblock {\em Ecology Let.}, 7:60--68, 2004.

\bibitem{TaNaBasic}
Kim Christensen, Simone~A. di~Collobiano, Matt Hall, and Henrik~J. Jensen.
\newblock Tangled nature: A model of evolutionary ecology.
\newblock {\em J. Theor. Biol.}, 216:73--84, 2002.

\bibitem{TaNaNetwork}
Paul Anderson and Henrik~Jeldtoft Jensen.
\newblock Network properties, species abundance and evolution in a model of
  evolutionary ecology.
\newblock {\em Journal of Theoretical Biology}, 232:551--558, 2005.

\bibitem{TaNaEcoli}
Daniel Lawson, Henrik~Jeldtoft Jensen, and Kunihiko Kaneko.
\newblock Diversity as a product of interspecial interactions.
\newblock {\em arXiv:q-bio.PE}, (0505019), 2005.

\bibitem{Connor1979}
Edward~F. Connor and Earl~D. McCoy.
\newblock The statistics and biology of the species-area relationship.
\newblock {\em Am. Nat.}, 113:791--833, 1979.

\bibitem{RechtsteinerNeutralGspace}
Andreas Rechtsteiner and Mark~A. Bebau.
\newblock A generic neutral model for measuring excess evolutionary activity of
  genotypes.
\newblock {\em GECCO-99: Proceedings of the Genetic and Evolutionary
  Computation Conference, July 13-17, 1999, Orlando, Florida}, pages
  1366--1373, 1999.

\bibitem{TaNaTimeDep}
Matt Hall, Kim Christensen, Simone~A. di~Collobiano, and Henrik~Jeldtoft
  Jensen.
\newblock Time-dependent extinction rate and species abundance in a
  tangled-nature model of biological evolution.
\newblock {\em Phys. Rev. E}, 66, 2002.

\bibitem{TaNaQuasi}
Simone~Avogadro di~Collobiano, Kim Christensen, and Henrik~Jeldtoft Jensen.
\newblock The tangled nature model as an evolving quasi-species model.
\newblock {\em J. Phys A}, 36:883--891, 2003.

\bibitem{StaufferPatchFoodWebs}
Dietrich Stauffer, Ambarish Kunwar, and Denashish Chowdhury.
\newblock Evolutionary ecology in silico: evolving food webs, migrating
  population and speciation.
\newblock {\em Physica A}, pages 202--215, 2005.

\bibitem{FitnessLandscapes}
Sergey Gavrilets.
\newblock {\em Fitness Landscapes and the Origin of Species}.
\newblock Princeton University Press, 41 William Street, Princeton, New Jersey
  08540, 2004.

\bibitem{SimonsTaNa}
Simon Laird and Henrik~Jeldtoft Jensen.
\newblock The tangled nature model with inheritance and constraint:
  Evolutionary ecology constricted by a conserved resource.
\newblock Unpublished work, 2005.

\bibitem{LargeSpeciesShifts}
Egbert~H. van Nes and Marten Scheffer.
\newblock Large species shifts triggered by small forces.
\newblock {\em Am. Nat.}, 164:255--266, 2004.

\bibitem{TokitaStableNetwork}
Kei Tokita and Ayumu Yasutomi.
\newblock Emergence of a complex and stable network in a model ecosystem with
  extinction and mutation.
\newblock {\em Theor. Pop. Biol.}, pages 131--146, 2003.

\bibitem{MaynardSmithGT}
J.~Maynard Smith.
\newblock {\em Evolution and the theory of games}.
\newblock Cambridge University Press, The Edinburgh Building, Cambridge CB2
  2RU, 1982.

\bibitem{MeasBioDiv}
Anne~E. Magurran.
\newblock {\em Measuring Biological Diversity}.
\newblock Blackwell Science Limited, 108 Cowley Road, Oxford, OX4 1JF, 2004.

\bibitem{RikvoldTanaBasic}
Per~Arne Rikvold and R.~K.~P. Zia.
\newblock Punctuated equilibria and 1/f noise in a biological coevolution model
  with individual-based dynamics.
\newblock {\em Phys. Rev. E}, 68, 2003.

\bibitem{RikvoldTaNaFluctuations}
R~K~P Zia and Per~Arne Rikvold.
\newblock Fluctuations and correlations in an individual-based model of
  evolution.
\newblock {\em J. Phys. A}, 37:5135--5155, 2004.

\bibitem{IslandBiogeog}
R.~H. MacArthur and E.~O. Wilson.
\newblock {\em The Theory of Island Biogeography}.
\newblock Princeton University Press, Princeton, NJ, USA, 1967.

\bibitem{MagurranNature}
Anne~E. Magurran and Peter~A. Henderson.
\newblock Explaining the excess of rare species in natural species abundance
  distributions.
\newblock {\em Nature}, pages 714--716, 2003.

\bibitem{DiverseEcoli}
Akiko Kashiwagi, Wataru Noumachi, Masato Katsuno, Mohammad~T. Alam, Itaru
  Urabe, and Tetsuya Yomo.
\newblock Plasticity of fitness and diversification process during an
  experimental molecular evolution.
\newblock {\em J. Molec. Evol.}, 52:502--509, 2001.

\bibitem{Biology}
N.~A. Cambell.
\newblock {\em Biology}.
\newblock Benjamin/Cummings Publ. Co., 2725 Sand Hill Road, Menlo Park,
  California 94025, 1996.

\bibitem{EvEcol}
Erik~R. Pianka.
\newblock {\em Evolutionary Ecology}.
\newblock Addison Wesley Educational Publishers, 1301 Sansome St, San
  Francisco, CA 94111, 2000.

\bibitem{DrosselReview}
Barbara Drossel.
\newblock Biological evolution and statistical physics.
\newblock {\em Cond. Mat.}, 2001.

\bibitem{HubbellRicklefs}
Stephen~P. Hubbell.
\newblock Modes of speciation and the lifespans of species under neutrality: a
  response to the comment of robert e. ricklefs.
\newblock {\em OIKOS}, 100:193--199, 2003.

\bibitem{RicklefsNeutrality}
Robert~E. Ricklefs.
\newblock A comment on hubbell's zero-sum ecological drift model.
\newblock {\em OIKOS}, 100:185--192, 2003.

\bibitem{PurvesPacala}
Drew~W. Purves and Stephen~W Pacala.
\newblock Ecological drift in niche-structured communities: neutral pattern
  does not imply neutral process.
\newblock In D.~Burslem, M.~Pinard, and S.~Hartley, editors, {\em Biotic
  Interactions in the Tropics - Their Role in the Maintenance of Species
  Diversity}. Cambridge University Press, The Edinburgh Building, Shaftesbury
  Road, Cambridge, CB2 2RU, 2005.

\end{thebibliography}

\end{scriptsize}
\end{spacing}
\end{document}